\documentclass[preprint,12pt]{elsarticle}
\usepackage{amsmath}
\usepackage{amssymb}
\usepackage{color}
\usepackage{epstopdf}

\begin{document}

\begin{frontmatter}

\title{Central governance based on monitoring and reporting solves the collective-risk social dilemma}

\author[chn,he]{Nanrong He}
\author[chn]{Xiaojie Chen\corref{correspondingauthor}}
\cortext[correspondingauthor]{Corresponding author}
\ead{xiaojiechen@uestc.edu.cn}
\author[hun]{Attila Szolnoki}
\address[chn]{School of Mathematical Sciences, University of Electronic Science and Technology of China, Chengdu 611731, China}
\address[he]{College of Science, Guilin University of Aerospace Technology, Guilin 541004, China}
\address[hun]{Institute of Technical Physics and Materials Science, Centre for Energy Research, Hungarian Academy of Sciences, P.O. Box 49, H-1525 Budapest, Hungary}

\begin{abstract}
Monitoring and reporting incorrect acts are pervasive for maintaining human cooperation, but in theory it is unclear how they influence each other. To explore their possible interactions we consider spatially structured population where individuals face the collective-risk social dilemma. In our minimal model cooperator players report defection according to the loss of their interests. In parallel we assume a monitoring institution that monitors all group member and identifies wrong behavior with a certain probability. In response to these feedbacks a sanctioning institution develops punishment schemes by imposing fines on related defector players stochastically. By means of Monte Carlo simulations, we find that the introduction of monitoring and reporting mechanisms can greatly promote the evolution of cooperation and there exists a sudden change of the cooperation level by varying model parameters, which can lead to an outbreak of cooperation for solving the collective-risk social dilemma.
\end{abstract}

\begin{keyword}
\texttt cooperation\sep\ punishment\sep\ monitoring\sep\ reporting\sep\ common resource
\end{keyword}

\end{frontmatter}

\section{Introduction}
\indent Ensuring sufficiently high level of cooperation and maintaining sustainable use of common resources are essential tasks for human societies \cite{ostrom_90,poteete_10,young_89}. However, selfish behavior often threatens public cooperation since it can provide a higher individual income \cite{du_wb_epl09,zhang_jtb13,zhang_sr15,liu_epl17,guo_amc17,chen_ploscb18,zhang_amc18,li_pnas18,wang_nc18}.
Bowing to such temptation of short-term interests leads to an excessive use of common resources and ultimately ``the tragedy of the commons'' state seems to be unavoidable \cite{hardin_g_s68}. Correspondingly, different ecological crisis \cite{plumwood_02} is identified, including the emergence of ozone hole \cite{solomon_rg99}, environmental pollution \cite{holdgate_80}, soil erosion \cite{shi_jae00}, and climate warming \cite{du_jm_epl14, pacheco_plrev14}. Stripped of particularities, these problems are identified as a sort of collective-risk social dilemma \cite{milinski_pnas08}, in which the declared collective target is in jeopardy, which has serious long-term consequences \cite{greenwood_epl11, chen_xj_epl12, chen_xj_pre12b}.

Theoretical and experimental studies have proposed effective means to promote public cooperation in the collective-risk social dilemma game \cite{perc_pr17}, such as reward \cite{szolnoki_epl10,rezaei_ei09, sasaki_jtb11, szolnoki_njp12}, punishment \cite{fehr_n02, boyd_pnas03, brandt_prsb03, helbing_ploscb10,wang_sr13,gao_l_srep15,yang_amc18,wang_amc18}, and exclusion \cite{sasaki_prsb13, li_k_pre15, szolnoki_pre17, liu_lj_srep17}. In parallel, some other control mechanisms are also identified in everyday life. For example, the monitoring-based and reporting-based governance has been used for the forest commons management \cite{rustagi_s10,yang_pnas13}, which is found to hinder the free-riding problem. In general, the top-down governance of monitoring and reporting means that once the feedbacks of monitoring and reporting are received, the administrative bureau will punish the defector based on available information. Such governance schemes are particularly common for public resources in human society. More precisely, a monitoring department monitors wrongdoers by random checks and provides information to a superior authority, while customers also report their offenders according to their own degree of damage. As a result, the administrative department imposes fines on the related enterprises based on the mentioned feedback information. This management system is pervasive in our controlled society and serves as a key control mechanism to maintain the quality of products high.

\indent Motivated by these practical examples, in this work we propose a game-theoretical model which considers the simultaneous presence of monitoring and reporting mechanisms to explore their collective impact on public cooperation. We assume that when a selfish behavior is observed, the involved cooperative players report defection to the external sanction institution based on the extent of loss of their income. In parallel, a top-down organized institution monitors players and reveals defectors from time to time. In combination with the feedback information from the monitoring institution and individual reporting behaviors, a sanction institution may enforce the corresponding punishment and accordingly imposes a fine on the related defectors.

We integrate these assumptions about monitoring and reporting into a previously studied model of collective-risk social dilemma \cite{chen_xj_srep14}, in which cooperative players contribute a part of its endowment to refill the collective target, while defective players retain all the endowment for themselves. Without external control mechanisms, it is found that in the collective-risk dilemma model excessive abundant common resource deters cooperative behavior and inefficient allocation of resources leads to the collapse of cooperation \cite{chen_xj_srep14}. Interestingly, we find that when monitoring and reporting are considered, the evolution of cooperation can be greatly promoted even if the allocation endowment for individuals is small. We further find that there exists a sudden change of the cooperation level when varying the model parameters, which can lead to an outbreak of cooperation for solving the collective-risk social dilemma.

\section{Model}

\indent We consider a population of individuals who play a public goods game on a $L\times L$ square lattice with periodic boundary conditions \cite{szolnoki_pre09c,perc_jrsi13}. Each individual $x$ forms a five-member group with nearest neighbors, hence every player takes part in five overlapping groups and the group size $G$ is five. Initially a player is designated as a cooperator or a defector with equal probability. At time step $t$, the endowment $a_{x}^{i}(t)$ for player $x$ in the group $i$ is defined as \cite{chen_xj_srep14}
\begin{eqnarray}
\ a_{x}^{i}(t)=\begin{cases}
 b, &\textrm{if} \hspace{2mm} R^{i}(t)\geq Gb\\
 R^{i}(t)/G, &\textrm{if} \hspace{2mm} R^{i}(t)< Gb
\end{cases}.
\end{eqnarray}
Here $R^{i}(t)$ is the amount of common resource available to the related group $i$ and $b$ represents the maximal possible endowment assigned to a player. Due to the overlapping groups the total income of the player on site $x$ can be obtained from five sources, given as $T_{x}(t)=\sum_{i=1} ^{i=5} a_{x}^{i}(t)$. Meanwhile cooperators contribute a fixed amount $c$ to the common pool in order to prevent the depletion, while defectors contribute nothing.

\indent Subsequently, cooperators may take individual reporting action based on the comparison between the total income from the common pools and the total cost they paid. Specifically, at time step $t$ player $x$ will report the free-riding behavior around itself to a top-down organized sanctioning institution when $T_{x}(t)<Gc$. In other words, the reporting probability $\rho_{x}(t)=1$ in this case. Otherwise, a cooperator reports the free-riding behavior around itself with a probability depending on the income difference given as
\begin{equation}
\ \rho_{x}(t)=1/\{1+\exp[(T_{x}(t)-T_{x}(t-1))/K_{r}]\},
\end{equation}
where $K_{r}$ characterizes the uncertainty of reporting action and $T_{x}(t-1)$ is the total income of player $x$ at time step $t-1$. Without losing generality we use $K_{r}=0.5$ implying that a cooperator prefers reporting bad behavior when the income is decreased, but there is still a minor chance to report when the income is increased. Notably, when cooperator $x$ undertakes the reporting action in the group, it should pay a reporting cost $\varepsilon$. Accordingly, the payoff of cooperator $x$ from group $i$ is thus $I_{x}^{i}(t)= a_{x}^{i}(t)-c-\varepsilon$. Otherwise, its payoff from group $i$ is $I_{x}^{i}(t)= a_{x}^{i}(t)-c$. Besides, we assume that the reporting probability of defectors is zero. And we assume that the total income of cooperators is zero before the first time step since there are no game interactions. On the other hand, the top-down monitoring mechanism is used for inspecting individual behavior. For simplicity, we assume that the group's behaviors are monitored with a fixed probability $\rho_M$ ($0\leq \rho_M\leq 1$) \cite{chen_sr15}. Thus defection behavior is detected with the probability $\rho_M$.

\indent As we already stressed, the punishment schemes of a top-down organized sanctioning institution are based on the information collected from monitoring and reporting actions. More specifically, defectors (if present) in the group $i$ will be punished with a probability which is a weighted mean of the monitoring probability and the average reporting probability of cooperators given as
\begin{eqnarray}
P^{i}(t)=(1-\omega)\rho_{R}^{i}(t)+\omega\rho_{M},
\end{eqnarray}
where $\rho_{R}^{i}(t)=\sum_{x\in i}\rho_{x}(t)/N_{C}^{i}$ is the average reporting probability of the cooperators in the group $i$, $N_{C}^{i}$ is the number of cooperators in the group, and $\omega$ is a parameter characterizing the relative weight between the reporting probability and the monitoring probability ($0\leq \omega\leq1$). When defector $x$ is punished by the sanctioning institution, its payoff is reduced by a fine $\lambda$, hence $I_{x}^{i}(t)=a_{x}^{i}(t)-\lambda$, otherwise its payoff collected from group $i$ is $I_{x}^{i}(t)=a_{x}^{i}(t)$.

\indent Because of the overlapping groups, the total payoff $I_{x}(t)$ of player $x$ is simply the sum over all related $I_{x}^{i}(t)$ payoff values from the overlapping groups. Notably, in our model the monitoring cost and the punishment cost are not considered explicitly for individuals in the game since they are covered by the top-down organized external institution.

\indent Starting with $R^{i}(0)=R_{0}$ in all groups, after individuals playing the games the updating protocol of the amount of common resources in each group is defined as
\begin{eqnarray}
R^{i}(t)=R^{i}(t-1)+\sum_{x\in i}[\alpha s_{x}c-a_{x}^{i}(t)],
\end{eqnarray}
where $R^{i}(t)$ is the amount of common resource (public goods) available to the group $i$ at time step $t$ and $\alpha$ is the synergy factor to the amount of contribution \cite{chen_xj_srep14}. For simplicity, we set $c=1$ in this study.

\indent After playing the games in all related groups player $x$ adopts the strategy of a randomly chosen neighbor $y$ with the probability
\begin{eqnarray}
q=\frac{1}{1+\exp\{[I_{x}(t)-I_{y}(t)]/K_{s}\}},
\end{eqnarray}
where $K_{s}$ denotes the amplitude of noise for strategy updating \cite{szabo_pre98}. Without losing generality we set $K_{s}=0.5$. This value ensures that it is very likely that a better performing player will pass its strategy to their neighbors, yet it is also possible that a player will occasionally learn from a less successful neighbor.

In our simulations, the key quantity for characterizing the cooperative behavior of the system is the density of cooperators, which is defined as the fraction of cooperators in the whole population. In our study synchronous updating protocol is applied. After a suitable transient time, the system evolves into a dynamical equilibrium state and the cooperator density reaches its asymptotic value where the fluctuations remain as small as $0.003$.
All the simulations were carried out on a square lattice with the size between $L=100$ and $L=2000$, where initially the two strategies of cooperators and defectors are randomly distributed among the population with an equal probability $0.5$.

\section{Results}

\begin{figure}
\centering
\includegraphics[width=15.0cm]{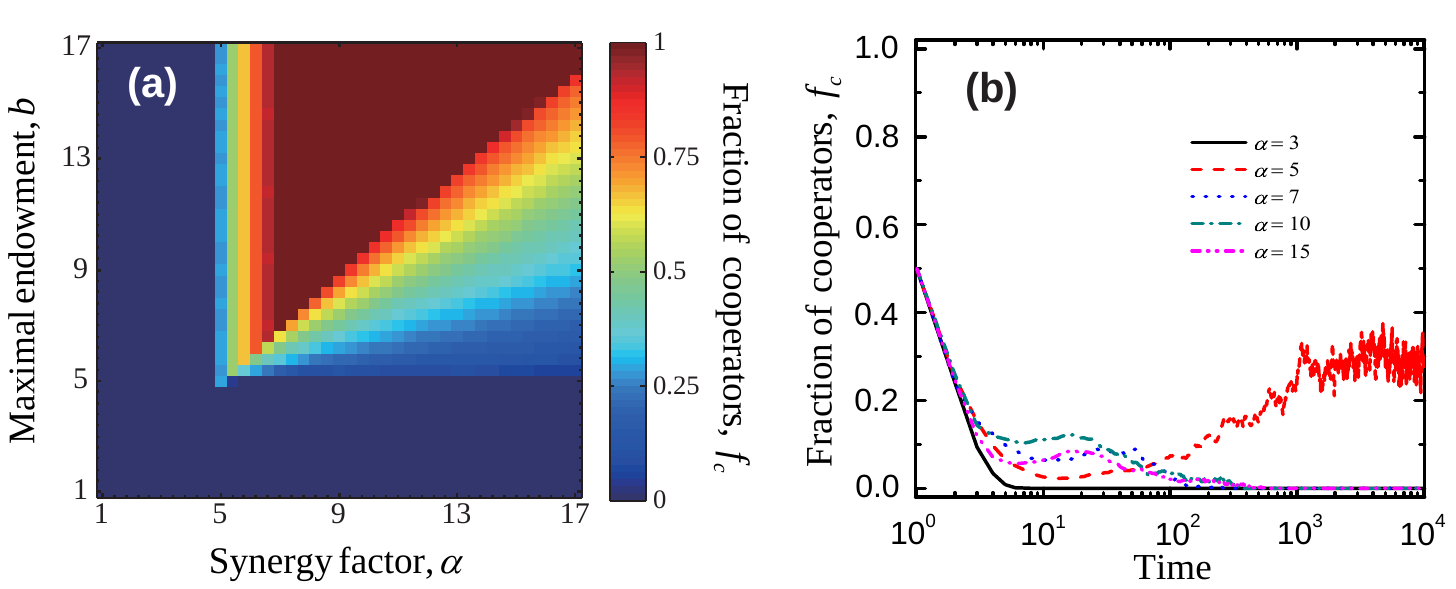}\\
\caption{Panel~(a) shows the stationary fraction of cooperators $f_{c}$ in dependence on the maximal possible endowment $b$ and the synergy factor $\alpha$ in the original collective-risk social dilemma model. Panel~(b) further depicts the fraction of cooperators $f_{c}$ as the function of time for five different values of $\alpha$ at $b=5$ in the case without the reporting-based and monitoring-based governance scheme. Other parameters are $R_{0}=25$ and $L=100$.}\label{F1}
\end{figure}

\begin{figure}
\centering
\includegraphics[width=13.5cm]{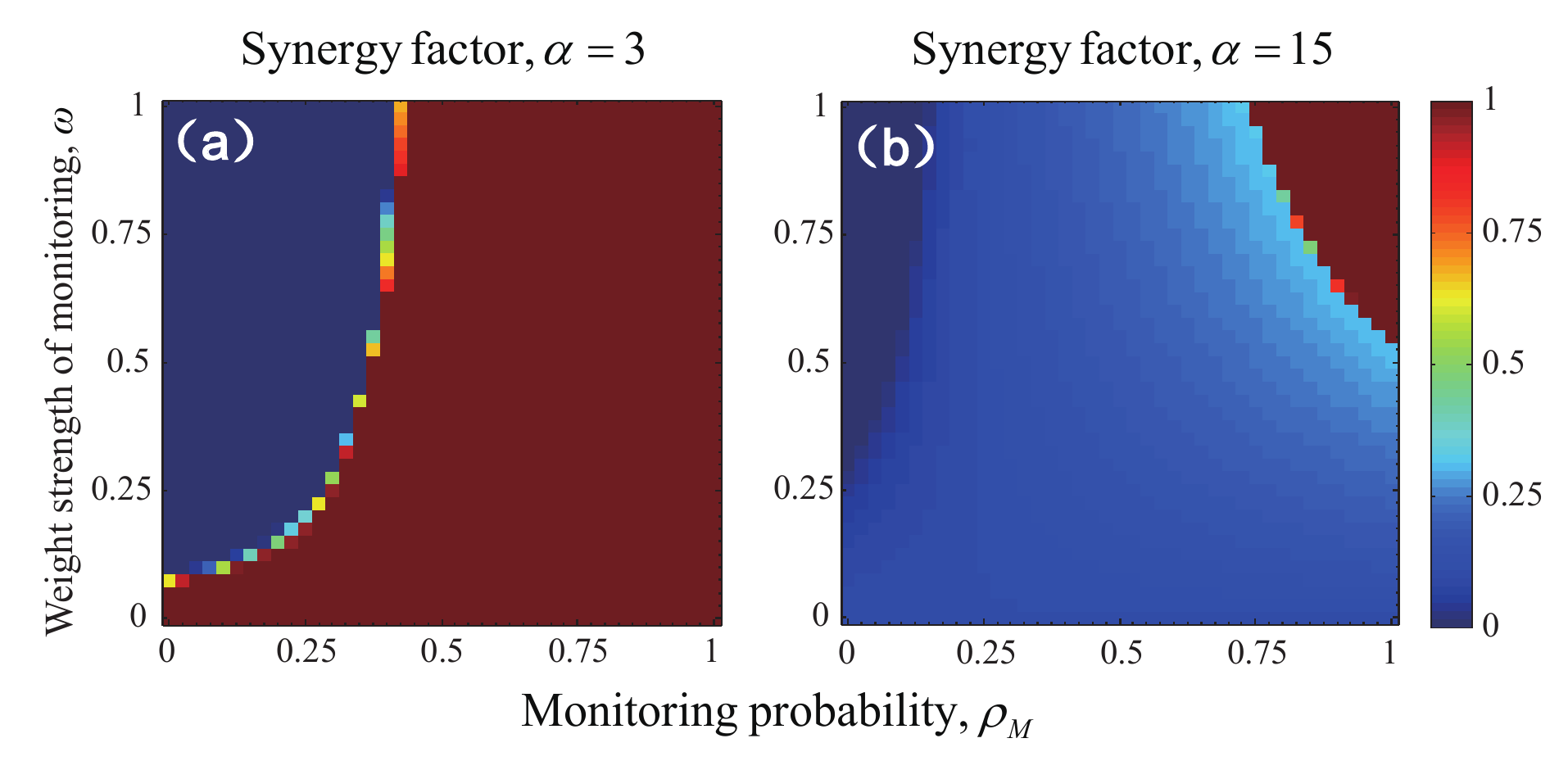}\\
\caption{Panels~(a) and (b) depict the fraction of cooperators $f_{c}$ in dependence on monitoring probability $\rho_{M}$ and weight strength of monitoring $\omega$ for $\alpha=3$ and $\alpha=15$ respectively. Other parameters are $b=5$, $\lambda=1.6$, $\varepsilon=0.5$, and $R_{0}=25$.}\label{F2}
\end{figure}

\indent For the sake of comparison, we first briefly summarize the evolutionary outcomes when the monitoring and reporting mechanisms are not considered in the proposed collective-risk social dilemma game \cite{chen_xj_srep14}. In Fig~\ref{F1}(a) we present the stationary fraction of cooperators $f_{c}$ in dependence on the maximal possible endowment $b$ and the synergy factor $\alpha$. We observe that the larger the values of the max endowment $b$, the broader the intermediate interval of the synergy factor $\alpha$ where high level of cooperation can be reached. This suggests that the common resources should be allocated rather than restricted and cooperation cannot be effectively maintained for a small or large value of $\alpha$ \cite{chen_xj_srep14}. To further illustrate this finding we show how the cooperation level evolves at low $b$ values as shown in Fig.~\ref{F1}(b). It demonstrates clearly that only adverse condition of $\alpha=5$ can ensure a limited level of cooperation $f_{c}\approx0.29$. But large or small values of $\alpha$ will always cause the collapse of cooperation. These results indicate that in this collectiver-risk social dilemma model a harsh condition is created for the evolution of cooperation when the synergy factor is too small or too large for small endowment $b$ values. We are then interested in investigating whether the introduction of monitoring and reporting mechanisms can promote the evolution of cooperation in such harsh conditions.

\indent As an answer, in Fig.~\ref{F2} we thus respectively show the fraction of cooperators for a small synergy factor value and for a large synergy factor value  in the case where both monitoring and reporting mechanisms are considered. Specifically, Fig.~\ref{F2} depicts the fraction of cooperators in dependence on the monitoring probability $\rho_M$ and the weight strength of monitoring to punishment $\omega$. We find that a nonzero cooperation level can be ensured no matter whether the synergy factor $\alpha$ is small [panel (a)] or large [panel (b)]. In addition, the region for full cooperation level for $\alpha=3$ is much larger than that for $\alpha=15$. These results indicate that the introduction of monitoring-based and reporting-based governance scheme provides a significant improvement for the evolution of cooperation and the governance scheme can work better for the evolution of cooperation at a small synergy factor value than at a large synergy factor value.

\begin{figure}
\centering
\includegraphics[width=13.5cm]{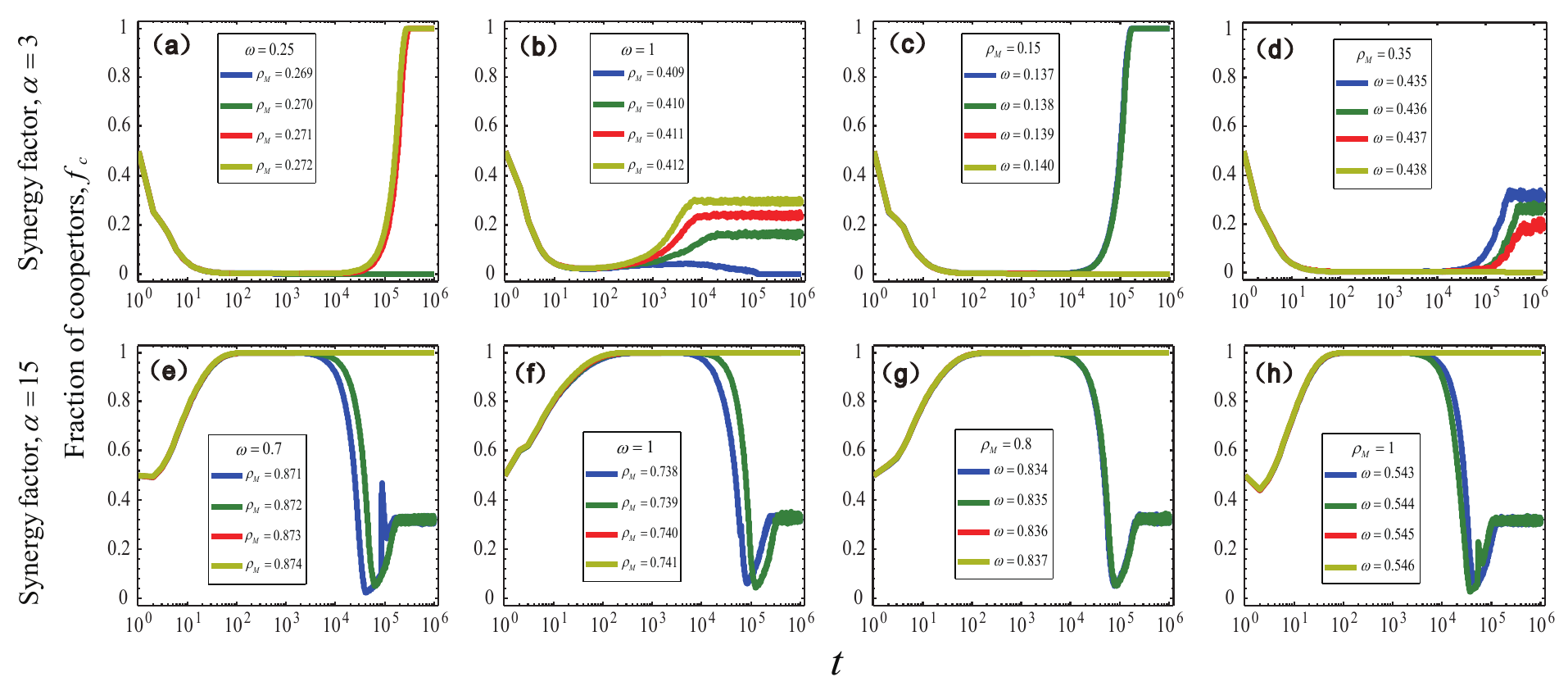}\\
\caption{Top row depicts the time evolution of cooperation for small $\alpha=3$. Panels~(a) and (b) depict the fraction of cooperators as a function of time for different values of $\rho_M$ at fixed $\omega=0.25$ and $\omega=1$ respectively. Panels~(c) and (d) depict the fraction of cooperators as a function of time for different value of $\omega$ at fixed $\rho_M=0.15$ and $\rho_M=0.35$ respectively. Bottom row depicts the time evolution of cooperation for large $\alpha=15$. Panels~(e) and (f) depict the fraction of cooperators as a function of time for different values of $\rho_{M}$ at fixed $\omega=0.7$ and $\omega=1$ respectively. Panels~(g) and (h) depict the fraction of cooperators as a function of time for different values of $\omega$ at fixed $\rho_{M}=0.8$ and $\rho_{M}=1$ respectively. Other parameters are $b=5$, $\lambda=1.6$, $\varepsilon=0.5$, and $R_{0}=25$.}\label{F3}
\end{figure}

\begin{figure}
\centering
\includegraphics[width=13.5cm]{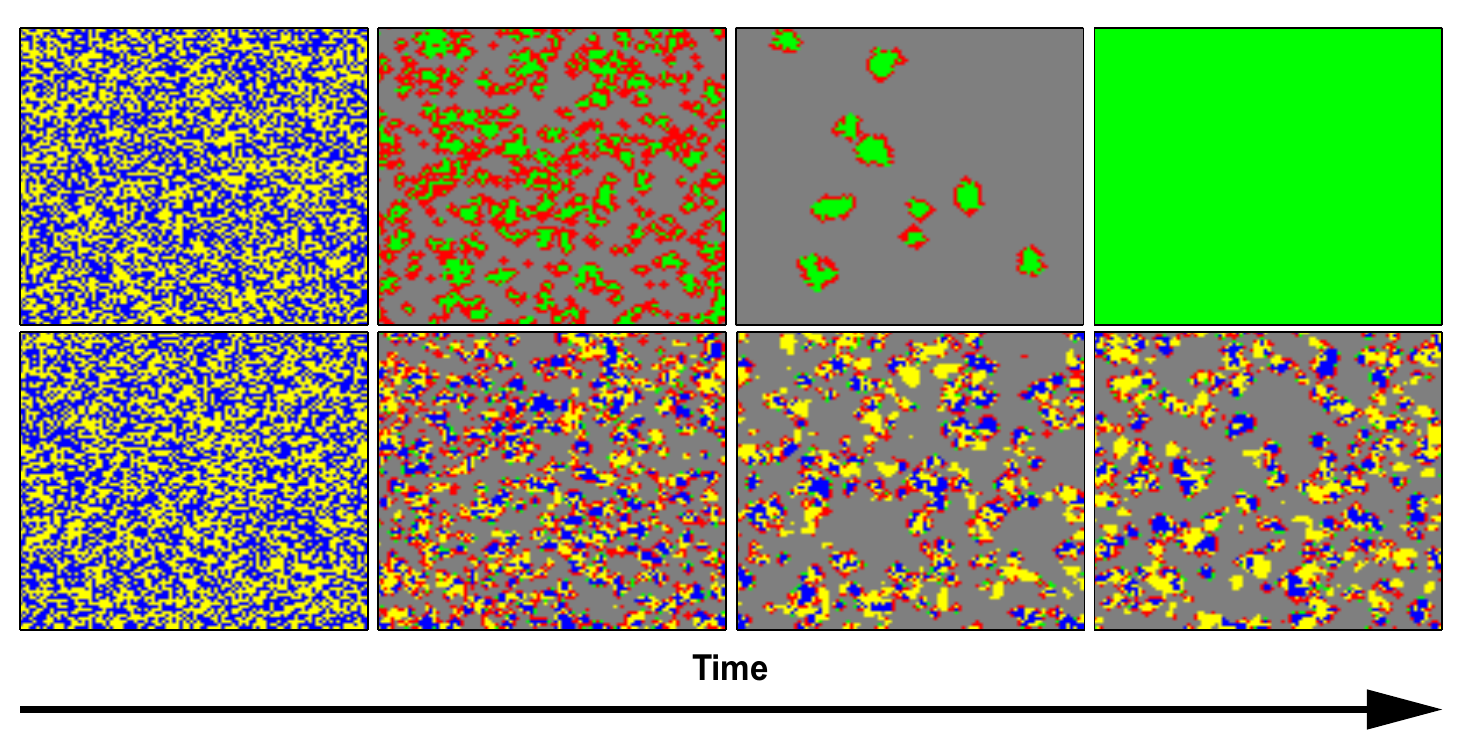}\\
\caption{Top row denotes the time evolution of spatial patterns for $\alpha=3$. Bottom row denotes the time evolution of spatial patterns for $\alpha=15$. Other parameters are $\rho_{M}=0$, $\omega=0$, $b=5$, $\lambda=1.6$, $\varepsilon=0.5$, and $R_{0}=25 $.}\label{F4}
\end{figure}

\indent We further unveil the effects of monitoring probability and weight strength of monitoring on the evolution of cooperation in Fig.~\ref{F2}. On one hand, we can see that for large values of $\omega$, the cooperation level increases monotonically by increasing the value of $\rho_{M}$. However, for the extremely low values of $\omega$ ($\omega\leq0.05$), the fraction of cooperators is almost unchanged as increasing the value of $\rho_{M}$. On the other hand, we can observe that when the monitoring probability $\rho_{M}$ is small, the fraction of cooperators decreases as the value of $\omega$ increases both for $\alpha=3$ and $\alpha=15$. However, for large values of $\rho_{M}$, the results are different. Specifically, for $\alpha=3$ cooperators become completely dominant when $\rho_{M}\geq0.48$. While for $\alpha=15$ the fraction of cooperators increases monotonically with the increase of $\omega$. In particular, the fraction of cooperators completely reaches its maximum $f_{c}=1$ only when $\omega$ and $\rho_{M}$ are both relatively large. Notably, we find that there exists an outbreak of cooperation when varying the monitoring probability value and the weight strength of monitoring, which implies discontinuous change of the cooperation level when varying these parameters.

\indent Furthermore, in order to illustrate how the cooperation level changes specifically with varying the monitoring probability or the weight strength of monitoring, we show the time evolution of the cooperation level for different values of $\rho_M$ and $\omega$ in Fig.~\ref{F3}. For $\alpha=3$ (top row of Fig.~\ref{F3}), we can see that the fraction of cooperators $f_{c}$ at the stationary state increases suddenly from zero to one as $\rho_{M}$ increases very slowly ($\omega$ decreases very slowly) for the small value $\omega=0.25$ ($\rho_{M}=0.15$), as illustrated in Fig.~\ref{F3}(a)[Fig.~\ref{F3}(c)]. However, for the large value $\omega=1$ ($\rho_{M}=0.35$), the fraction of cooperators at the stationary state increases gradually by increasing $\rho_{M}$ very slowly (decreasing $\omega$ very slowly), as illustrated in Fig.~\ref{F3}(b) [Fig.~\ref{F3}(d)]. While for $\alpha=15$ (bottom row of Fig.~\ref{F3}), we find that there also exist discontinuous changes where full cooperation state can be reached suddenly for specific values of $\omega$ and $\rho_{M}$. Taken together, our results show that adjusting properly the monitoring intensity and the weight strength of monitoring can lead to an outbreak of cooperation, where the evolutionary advantages of defectors over cooperators can be completely subverted in the system.

\begin{figure}
\centering
\includegraphics[width=13.5cm]{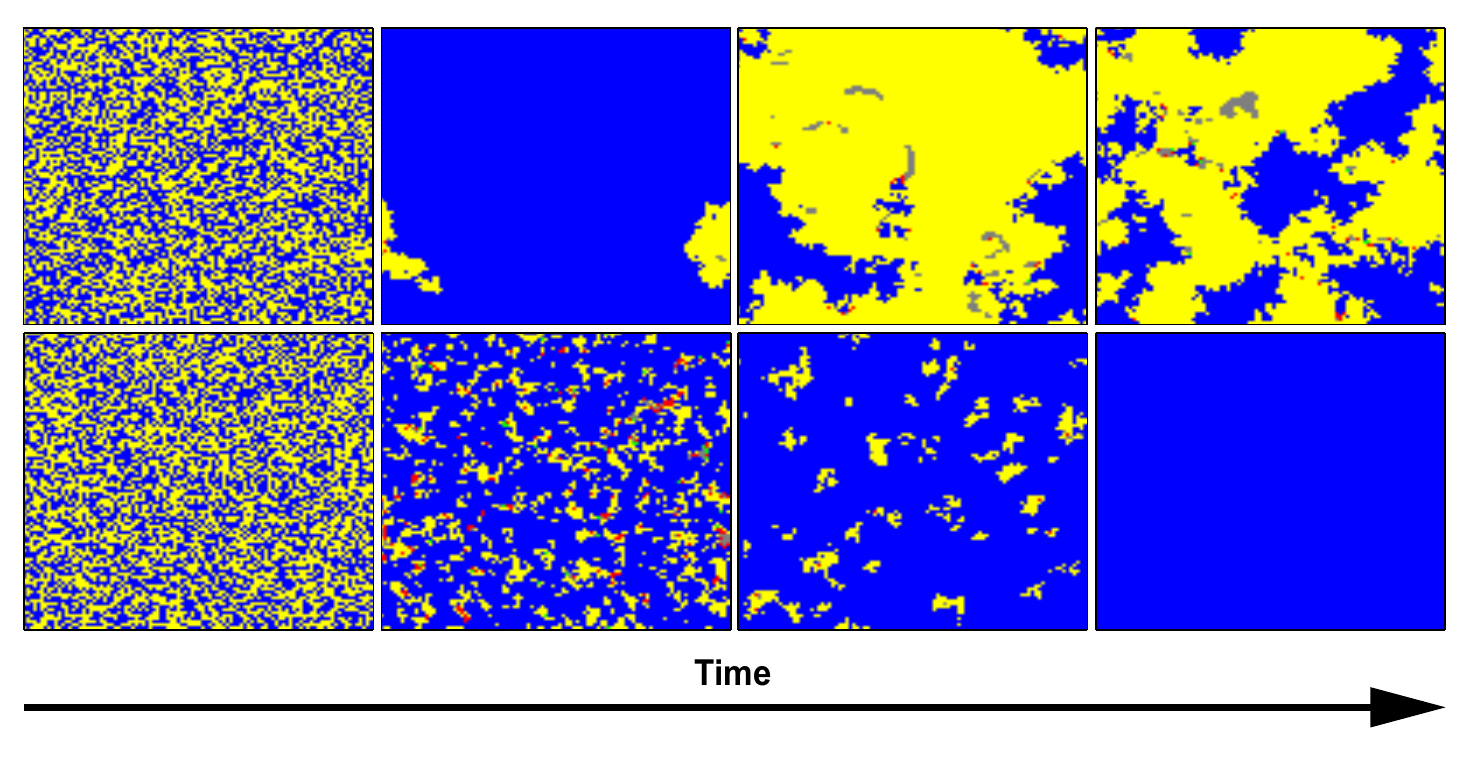}\\
\caption{Top row shows the time evolution of spatial patterns for $\rho_{M}=0.739$. Bottom row shows the time evolution of spatial patterns for $\rho_{M}=0.74$. Other parameters are $\alpha=15$, $\omega=1$, $b=5$, $\lambda=1.6$, $\varepsilon=0.5$, and $R_{0}=25 $.}\label{F5}
\end{figure}

\begin{figure}
\centering
\includegraphics[width=13.5cm]{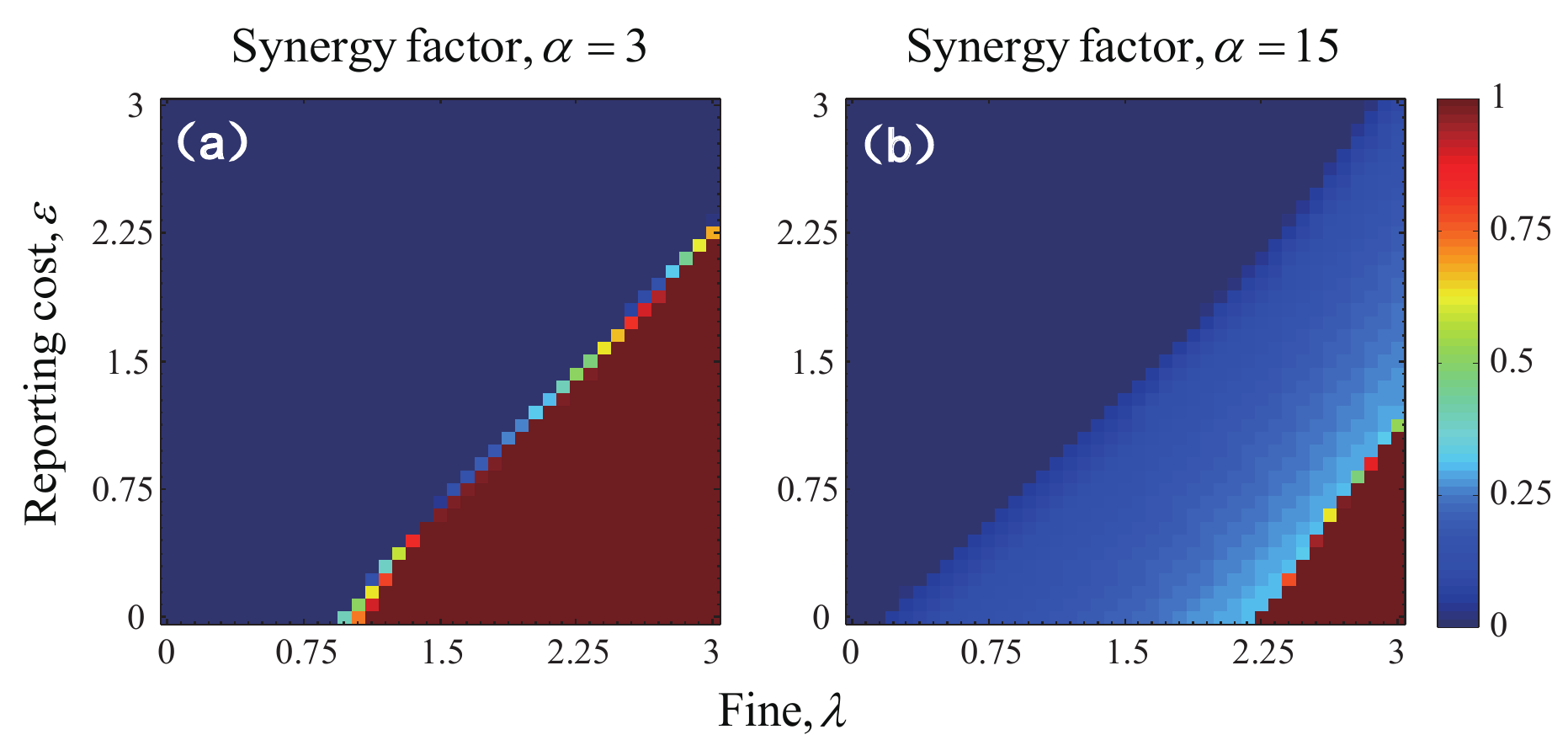}\\
\caption{The cooperation level $f_C$ in dependence on punishment fine $\lambda$ and reporting cost $\varepsilon$ for $\alpha=3$ [panel~(a)] and $\alpha=15$ [panel~(b)]. Other parameters are $\rho_{M}=0.5$, $\omega=0.5$, $b=5$, and $R_{0}=25$.}\label{F6}
\end{figure}

\indent To better understand our observations we present a series of snapshots of strategy distributions for some representative values of model parameters. For a deeper insight we use different colors not just for cooperator and defector strategies, but also depending on the available amount of common resources. In particular, blue (yellow) color represents cooperators (defectors) centered in a group where $R^{i}(t)\geq Gb$. Furthermore, green (red) color denotes cooperator (defector) player centered in a group where $R^{i}(t)<Gb$. Last, grey color marks defectors who are centered in a group where there are no common resources left (note that if cooperators are present the common resource is always larger than zero).

\indent We first present two representative evolutionary outcomes for low and high $\alpha$ values respectively. For $\alpha=3$ (top row of Fig.~\ref{F4}), cooperators can be exploited easily by defectors in the initial rounds, and thus the fraction of blue cooperators decrease rapidly. Only just a tiny portion of cooperators survives who manage to form a critical size of compact clusters. Then cooperators are more likely to destroy the defectors by raising their reporting vigilance and forming reporting clusters. As a result, they produce a relatively higher group benefit, which is marked by green color. Interestingly, red defectors who enjoy the relatively high product of cooperative groups can protect the group of grey defectors who have nothing. But sometimes this protective shield becomes leaky which offers a chance for green cooperators to propagate. Eventually, the cooperative clusters slowly expand and gradually dominate the entire population. This observation supports the general expectation that harsh environment could be useful for the community to select cooperation strategy as the only evolutionary escape route \cite{szolnoki_epl14}.
While for $\alpha=15$ (bottom row of Fig.~\ref{F4}), the situation is very different. The effectiveness of cooperators is so high that even a few cooperators in the group are sufficient to provide enough common resources for each group member. Thus, blue cooperators in the abundant common resources will get the maximum endowment. Consequently, they relax their vigilance, do not form a relatively larger clusters, and reduce their reporting intensity. This is why yellow defectors who refuse to contribute to the group can easily prevail against the blue domains of cooperators. Meanwhile, minor clusters of blue cooperators have an evolutionary advantage over red and grey defectors. Defectors and a small number of cooperators can finally coexist. This highlights that cooperators in the case of high $\alpha$ value are more likely to relax their vigilance, resulting in a low level of cooperation.

\indent We continue by illustrating the microscopic mechanism for a rapid switch of cooperation level when monitoring feedback is at work. To do that, we present series of representative snapshots for two different values of $\rho_{M}$ in Fig~\ref{F5}. For $\rho_{M}=0.739$ (top row of Fig.~\ref{F5}), it can be clearly seen that the sanction institution with higher monitoring probability can effectively attack the fragmented defectors, and thus cooperators quickly propagate in the entire population. Interestingly, some surviving defectors can form a certain type of clusters. When cooperators meet the clusters of defectors, the former needs to pay more reporting costs, which diminishes their evolutionary advantage and provides a surviving chance for defectors. Meanwhile, with the depletion of the common resources in the yellow domains, the scrappy cooperators around the grey domains can survive by defeating grey defectors. Thus such effect can lead to a stable coexistence of defector and cooperator players. Interestingly, when we change the value of $\rho_{M}$ just to $0.74$ (bottom row of Fig.~\ref{F5}), the evolutionary outcome is completely different. Here the scattered defectors are unable to resist the pressure of monitoring. They who have formed clusters attempt to penetrate into the blue area, but they cannot escape extinction. Fig.~\ref{F5} further illustrates that there exists a critical value of the monitoring intensity above which defectors cannot successfully invade cooperators' clusters and instead become extinct finally. When the monitoring intensity is below the critical value, some defectors can coexist with cooperator players. This thus explains clearly that there exists a sudden change of the cooperation level as the monitoring probability increases. Here we do not show the representative snapshots for different values of $\omega$, but we stress that similar spatial patterns to those in Fig.~\ref{F5} can be observed when varying the value of $\omega$.

\indent In what follows, it remains of interest to show how the fine $\lambda$ and reporting cost $\varepsilon$ influence the evolution of cooperation. Here, we present the corresponding stationary fraction of cooperators for $\alpha=3$ and $\alpha=15$, as illustrated respectively in Fig.~\ref{F6}~(a) and (b). As we can see from Fig.~\ref{F6}, reducing reporting costs $\varepsilon$ and raising fine value $\lambda$ are conducive to produce high levels of cooperation, even if the monitoring probability and decision propensity are relatively modest. This indicates that related institutions should try their best to reduce reporting costs and raise fines on defectors for promotion of cooperation. In addition, there still exists a rapid change of the cooperation level as the punishment fine or the reporting cost increases for the two different $\alpha$ values. This warrants that the appropriate level punishment fine should be adjusted carefully in order to reach the desired goal of high cooperation level \cite{helbing_njp10}.


\section{Discussion}
\indent We have introduced a monitoring- and reporting-based governance scheme into the collective-risk social dilemma game, and studied how their presence influences the evolution of cooperation in spatially structured populations. Motivated by the fact that these regulations are frequently applied in human societies for the governance of the common goods \cite{rustagi_s10,yang_pnas13}, we have integrated such top-down organized control mechanisms with punishment into a game-theoretical model. We emphasize that in our model when a defector is selected to be punished, cooperators do not incur a direct cost to defectors, whereas the top-down sanctioning institution imposes a fine on the defector. Thus the second-order free-riding problem does not appear in present setup and our principal aim is to investigate whether such top-down-type governance scheme can solve the dilemma of public cooperation. By means of Monte Carlo simulations, we find that the introduction of such governance scheme can significantly promote the evolution of cooperation even if the survival condition for cooperators in the original collective-risk social dilemma is extremely harsh.

\indent More interestingly, we find that there exist sudden changes of the cooperation level as the monitoring probability or the weight strength of monitoring to punishment are increased. If the monitoring probability or the weight strength of monitoring does not reach a critical threshold value required for full cooperation, cooperation will remain at a very basic level. On the other hand, only a tiny change is needed near the critical threshold, which will lead to a huge increase of cooperation level. From the perspective of the regulation about fines and reporting costs, there also exists a sharp transition from a low- to high-cooperative state. These observations suggest that the key point to reach a highly cooperative solution is to properly adjust the key parameters when the monitoring- and reporting-based governance schemes are used.

\section*{Acknowledgments}
This research was supported by the National Natural Science Foundation of China (Grant No. 61503062) and the Hungarian National Research Fund (Grant K-120785).

\bibliographystyle{elsarticle-num-names}

\end{document}